\documentclass[twocolumn,preprintnumbers,aps]{revtex4-1}
\usepackage{amsmath}
\usepackage{graphicx}
\usepackage{color}
\usepackage{subfigure}
\usepackage{hyperref}

\usepackage{siunitx}
\DeclareSIUnit{\atmosphere}{atm}

\newcommand{\BAN}{\ensuremath{B_{1g}\,}}
\newcommand{\BN}{\ensuremath{B_{2g}\,}}

\newcommand{\Ts}{\ensuremath{T^{\ast}\,}}
\newcommand{\Tc}{\ensuremath{T_{\rm c}\,}}
\newcommand{\cm}{\ensuremath{{\rm cm}^{-1}}}

\graphicspath{{images/}{../images/}}

\begin{document}
\title{Spin singlet and quasiparticles excitations in cuprate superconductors}
\author{M. Mezidi$^{1,2}$, A. Alekhin $^1$, G. D. Gu $^3$, D. Colson$^4$, S. Houver$^1$, M. Cazayous $^1$,Y. Gallais $^1$ and A. Sacuto$^{1*}$}
\affiliation{$^1$ Universit\'e Paris-cit\'e, Laboratoire Mat\'eriaux et Ph\'enom$\grave{e}$nes Quantiques, CNRS (UMR 7162), 75013 Paris, France\\
$^2$ D\'epartement de physique, Institut quantique and RQMP, Universit\'e de Sherbrooke, Sherbrooke, Qu\'ebec, Canada.\\
$^3$ Matter Physics and Materials Science, Brookhaven National Laboratory (BNL), Upton, NY 11973, USA.\\
$^4$ Universit\'e Paris-Saclay, CEA, CNRS, SPEC, 91191, Gif-sur-Yvette, France.}

\date{\today}

\begin{abstract}

We followed step by step the transition from an antiferromagnetic (AF) Mott insulator to a superconducting (SC) metal in the Bi$_2$Sr$_2$CaCu$_{2}$O$_{8+\delta}$ (Bi-2212) cuprate using the electronic Raman scattering spectroscopy. This was achieved by tracking the doping dependence of the spin singlet excitation originate from the AF Mott insulator,  the normal state quasiparticles excitation related to the mobile charge carriers and the Bogoliubov quasiparticles related to the SC gap. We show that the signature of the pseudogap phase which develops during this transition, can be interpreted as the blocking of charge carriers by the enhancement of the antiferromagnetic correlations as the temperature drops.  We find that the energy scale of the pseudogap, $\Delta_{\textrm{pg}}(p)$, closely follows the one of the spin singlet excitation, $\Delta_{\textrm{sse}}(p)$, with doping $p$. The quasiparticles lifetime considerably increases with doping when the pseudogap collapses. We reveal that the maximum amplitude of the SC gap, $\Delta_{\textrm{sc}}^{\textrm{max}}$ and the SC transition temperature \Tc  are linked in an extended range of doping such as $\Delta_{\textrm{sc}}^{\textrm{max}}(p) \propto \Delta_{\textrm{sse}}(p)\, \Tc(p)$. This relation suggests that the AF correlations play a key role in the mechanism of superconductivity. 

\end{abstract}

\maketitle

\section{I. Introduction}

The transformation from an antiferromagnetic (AF) Mott insulator to a metal superconductor as the holes number increases is a key element for understanding the physics of cuprates \cite{Anderson87,Scalapino1995,Orenstein00,Norman03,abanov03,Lee06,Kohsaka2008,Keimer2015}. In the AF Mott insulator phase, due to the strong one-site Coulomb repulsion $U$, electrons are localized on the copper sites and their spins are antiferromagnetically ordered along the Cu-O-Cu bonds in the copper oxygen planes due to the $J>0$ exchange interaction constant. When a small fraction of electrons are removed from the copper oxygen plane, what is known as the hole doping, $p$, the AF order is waning and the electrons start to move. As the temperature decreases below \Ts, an enigmatic phase emerges, the so-called pseudogap phase which harbors several electronic orders \cite{Keimer2015,Fauque06,Wu13, Daou2010,Ghiringhelli12,Loret2019,Proust2019,Frachet2020,Vinograd2022}. At lower temperature, below \Tc, the $d-$ wave superconducting (SC) phase settles down. At higher doping level, the pseudogap phase disappears, giving way to a strange metal and then to a Fermi liquid metal, with which superconductivity persists and finally disappears when the hole doping increases. The AF Mott insulator phase, the metallic phase and the superconducting phase are respectively characterized by the spin singlet excitation (SSE) stemming for the AF lattice, the quasiparticles excitation (QSPE) originate from the mobile charge carriers and the Bogoluibov quasiparticles excitation coming from the Cooper pairs breaking at twice the energy of the superconducting (SC) gap. All these three excitations are detectable by electronic Raman spectroscopy.\\
In the past, most of the Raman studies were focused on the spin singlet excitation (also called two-magnon) in cuprates. In particular, its relationship with the number of carriers or its interplay with the SC gap and its link to the pseudogap were investigated \cite{Lyons1988,Lyons1988a,Blumberg1997,Sugai2000,Sugai03,Tassini2008,Li2012,Chelwani2018,Wang2022}. However, these studies were severely limited by the range of doping (above $p=0.1$ and below $p=0.20$), which led to believe that first, the energy scale of the pseudogap phase, $\Delta_{\textrm{pg}}(p)$ and the one of the maximum amplitude of the superconducting gap, $\Delta_{\textrm {sc}}^{\textrm{max}}(p)$ are the same or very close to each other and secondly, that the energy scale of the spin singlet excitation, $\Delta_{\textrm{sse}}(p)$ and the one of $\Delta_{\textrm {sc}}^{\textrm{max}}(p)$ are proportional each other as a function of doping. \\ 
Here, we show by conducting an electronic Raman scattering study on Bi-2212 cuprate, over a larger range of doping from $p=0.05$ to $p=0.23$ and in an extended spectral range, that these two assertions are not valid over the full range of doping. We say that $\Delta_{\textrm{pg}}(p)$ scale is distinct from the one of $\Delta_{\textrm {sc}}^{\textrm{max}}(p)$ and it closely follows the one of $\Delta_{\textrm{sse}}(p)$.  By simultaneously tracking the doping dependence of the SSE and QSPE in the Raman spectra, we propose that the pseudogap opening as the temperature decreases is essentially due to the blocking of the charge carriers by the enhancement of the AF correlations that originate from the parent AF Mott insulator. This causes the loss of low energy quasiparticles spectral weight when the temperature drops, as originally described \cite{Alloul89,Norman1998,Timusk99,Tallon01}. We find that the quasiparticles lifetime exponentially increases with doping when approaching the pseudogap end. Finally, we show that the energy of the spin singlet excitation, $\Delta_{\textrm{sse}}$ and the maximum amplitude of the superconducting gap $\Delta_{\textrm {sc}}^{\textrm{max}}$ are not proportional to each other but connected to \Tc over a wide range of doping, such as: $\Delta_{\textrm {sc}}^{\textrm{max}}(p) \propto \Delta_{\textrm{sse}}(p) \, \Tc(p)$. We find this relation is consistent with earlier empirical laws extracted from other experimental techniques \cite{Ding01,Homes2005}. This relation suggests that the AF correlations have to be considered for understanding the cuprates superconducting state.

\section{II. Results and discussion}

\begin{figure}[ht]
\includegraphics[scale=0.55]{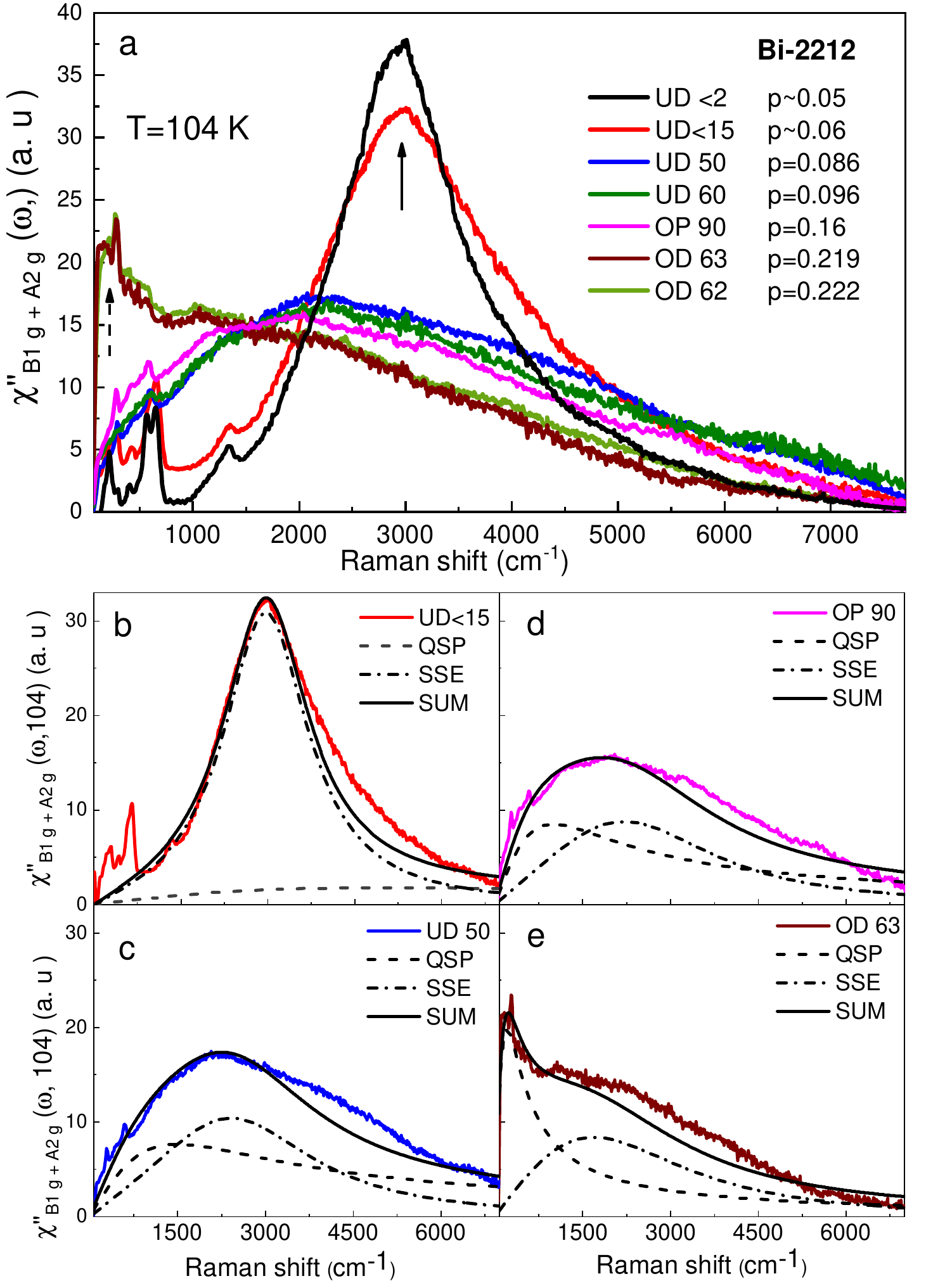}
\caption{(a) $\chi^{\prime \prime}_{B1g+A2g} (\omega)$ of Bi-2212 compound as a function of doping measured à T=$104$ K $>$ \Tc. The 532 nm excitation line was used. The initials UD, OP, OD stand for underdoped, optimally doped and overdoped. The numbers that follows the initials, correspond to the \Tc value. The doping level $p$, was deduced from $T_c$ using the Presland-Tallon s'law \cite{Presland91} and was corroborated by the energy of the pair breaking peak (see SI of \cite{benhabib15}). (b)-(d) Fits of the Raman responses for several doping levels which highlight the SSE contribution is plotted (dashed-dotted line) and the one of the QSPE in dashed line.}
\label{fig1}
\end{figure}

\begin{figure}[ht]
\includegraphics[scale=0.3]{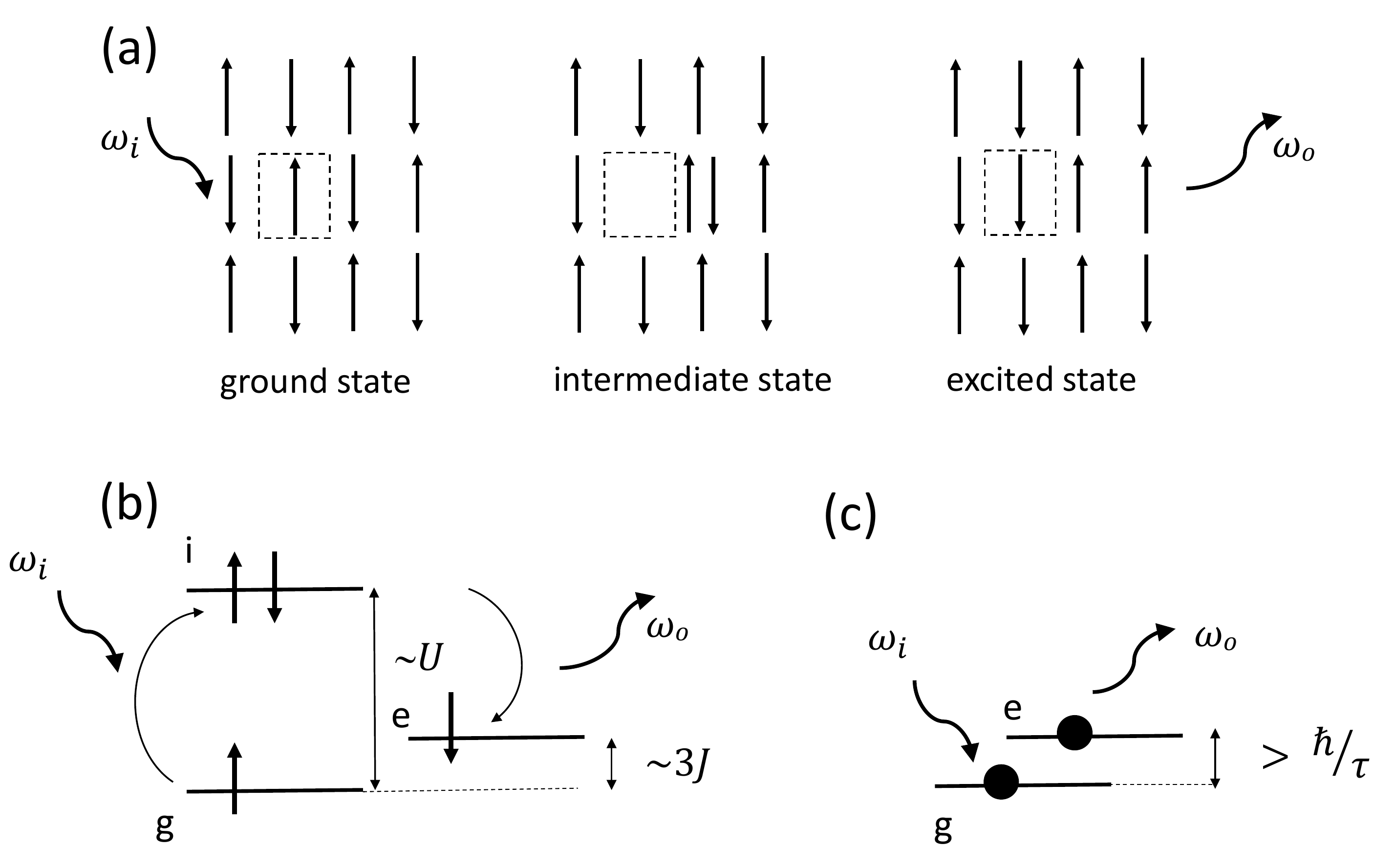}
\caption{(a) Spin singlet excitation in the real space of an AF lattice induced by inelastic light scattering. $\omega_{i,o}$ denotes respectively, the energy of the incoming and outgoing photon. Energy diagram of the inelastic light scattering process for (b) a spin singlet excitation in an insulator AF lattice and (c) a quasiparticle excitation in a metal. The letters g, i and e stand for the ground, intermediate and final state.} 
\label{fig2}
\end{figure}

We performed electronic Raman measurements on the (Bi-2212) system over a large spectral range.  This system has the advantage of allowing an exploration of its ($T-p$) phase diagram over a wide range of doping levels while only resorting to oxygen doping. Such a doping is much less invasive than cation insertion as done in YBCO or LSCO systems. The oxygen doped (Bi-2212) system has a maximum of \Tc=90 K. Details of the crystal growth and characterization are given in \cite{wen2008, benhabib15, Benhabib2015}. Electronic Raman scattering is a particularly useful probe for studying the cuprates properties \cite{LeTacon2006}. Indeed, it allows us to select distinct parts of the Brillouin zone namely, the anti-nodal and nodal regions (see Appendix A for details). In the \BAN geometry, the Raman form factor is $(\cos k_x -\cos k_y)^2$ and it predominantly probes the anti-nodal region where the superconducting gap and the pseudogap are maximal while in the \BN geometry the Raman form factor is $(\sin^2 k_x \sin^2 k_y)$ and it probes mostly the nodal region where the superconducting gap and the pseudogap are minimal. In this work, we focus on the \BAN Raman response function got from the ($B_{1g} + A_{2g}$) response (cf. Appendix A). It turns out that the $A_{2g}$ component is negligible in comparison to the \BAN one as it will be shown in a future article. The $\chi^{\prime \prime}_{B1g+A2g} (\omega, T=104 K)$ Raman response of Bi-2212 single crystals from the underdoped (UD) to the overdoped (OD) regime is reported in Fig.~\ref{fig1} (a).
%%%%%%%%%%%%%%%%%%%%%%%%%%%%%%%%%%%%%%%%%%%%%%%%%%%%%%%%%%%%%%%%%%%%%%%%%%%%%%%%%%%%%%%%%%%%%%%%%%%%%%%%%%%%%%%%%%
%%%%%%%%%%%%%%%%%%%%%%%%%%%%%%%%%%%%%%%%%%%%%%%%%%%%%%%%%%%%%%%%%%%%%%%%%%%%%%%%%%%%%%%%%%%%%%%%%%%%%%%%
We see, at low doping (0.05-0.06), a well-defined and intense peak centred around 3000 \cm (indicated by a full arrow in Fig.~\ref{fig1} (a)). It is assigned to the spin singlet excitation (SSE) related to the AF lattice. Its inelastic light scattering process is described in Fig.~\ref{fig2} (a) and (b). The incoming photon is absorbed by moving an up-spin to the neighbouring site occupied by a down-spin. The down-spin then takes the place of the formerly up-site emitting an outgoing photon \cite{Chubukov1995,Devereaux2007}. At the end of this process six singlets are destroyed in the AF lattice and this costs an effective energy $2\Delta_{\textrm{sse}}=J_{\textrm{eff}}\approx 3J$. $J$ is the exchange interaction constant. As the doping increases, the AF lattice breaks up and the magnitude of the effective exchange interaction $J_{\textrm{eff}}$ (initially equal to $3J$) decreases. The SSE mode is mainly detected in \BAN Raman response as predicted by the Fleury-Loudon  model for a $2D$, spin $\frac{1}{2}$ Heisenberg system with only nearest-neighbor interactions \cite{Fleury1968,Parkinson1969,Chubukov1995}. In the vicinity of the AF insulator phase, the SSE peak location has shown to give the first estimation of $J$ in lanthanum and yttrium-based cuprates \cite{Lyons1988a, Lyons1988}. In (UD$<$2K) Bi-2212,  we find $ J \approx$ 125 meV in good agreement with earlier works \cite{Sugai03}. The SSE  peak has a slightly asymmetrical profile in its high energy side  (Fig.~\ref{fig1} (a)). This was interpreted by extensions to the Loudon Fleury model \cite{Devereaux2007} including the triple resonant effect \cite{Chubukov1995} and  the magnon-magnon interactions mediated by the Higgs mode \cite{Weidinger2015} recently supported by Raman study \cite{Chelwani2018}. As the doping is increased, the SSE peak considerably broadens and its maximum shifts to low energy from 3000 \cm  to 1500 \cm (Fig.~\ref{fig1} (a)). In order to thoroughly follow the evolution of the SSE peak from the very under-doped regime to the moderately underdoped regime, we have reported in Appendix B (cf.  Fig.~\ref{fig5}) the Raman spectra of Bi-2212 in between $p\approx 0.06$ and $p\approx0.086$. We observe that the SSE peak softens in energy and broadens continuously as the doping increases. This is consistent with previous studies on other cuprates \cite{Sugai03,Tassini2008,Li2012}. Interestingly, when the doping reaches $p\approx 0.22$, a new peak emerges below $500$ \cm (indicated by a dashed arrow in Fig.~\ref{fig1} (a)). We assigned this peak to the quasiparticles excitation (QSPE) stemming from the mobile charge carriers. This peak is controlled by the scattering rate $\gamma\approx \frac{\hbar}{\tau}$ where $\tau$ is the quasiparticles lifetime. Its inelastic light scattering process is illustrated in Fig.~\ref{fig2} (c). The energy is transferred from the photons to the charge carriers which gain kinetic energy \cite{Shastry90,Devereaux2007}. In summary, as the doping increases, we observe that the high energy SSE peak softens in energy, broadens and weakens in intensity while the low energy QSPE peak grows up. 
%%%%%%%%%%%%%%%%%%%%%%%%%%%%%%%%%%%%%%%%%%%%%%%%%%%%%%%%%%%%%%%%%%%%%%%%%%%%%%%%%%%%%%%%%%%%%%%%%%%%%%%
%%%%%%%%%%%%%%%%%%%%%%%%%%%%%%%%%%%%%%%%%%%%%%%%%%%%%%%%%%%%%%%%%%%%%%%%%%%%%%%%%%%%%%%%%%%%%%%%%%%%%%%%

\begin{figure}[ht]
\includegraphics[scale=0.4]{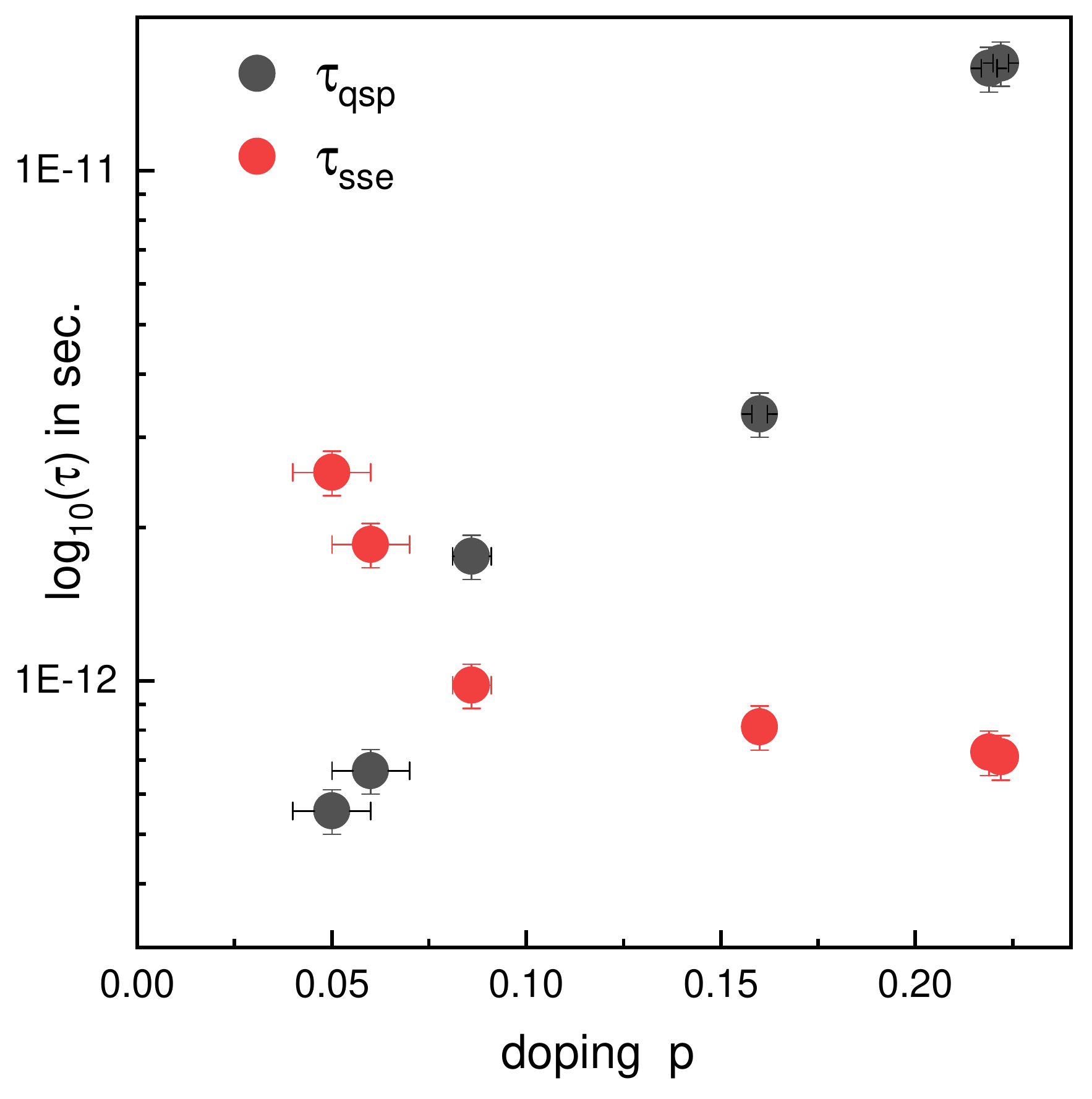}
\caption{Doping evolution of the spin singlet excitation and quasi-particle excitation lifetime. $\tau_{sse/qsp}=\frac{1}{c\gamma}_{sse/qsp}$ where $c$ is the light velocity.}
\label{fig3}
\end{figure}

\begin{figure*}[ht]
\includegraphics[scale=0.5]{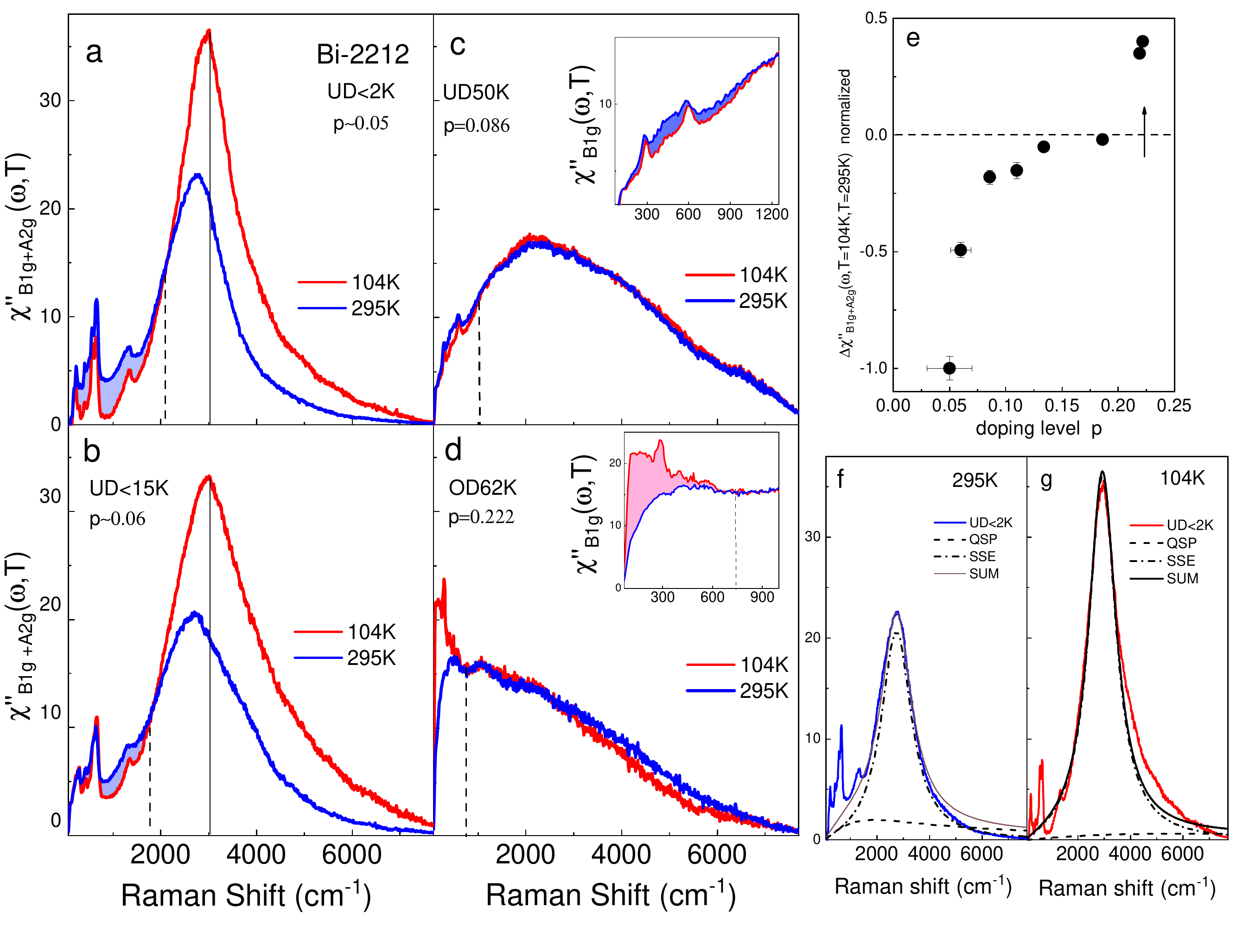}
\caption{(a) to (d) Raman response $\chi^{\prime \prime}_{B1g} (\omega)$ of Bi-2212 at T=295 K and 104 K for selected doping levels. The blue shaded area underlined the pseudogap spectral weight depletion as the temperature is lowered below the dotted line. The inset of panel (c) shows a zoom of the pseudogap depletion. The hardening of the SSE peak as $T$ decreases, is underlined by the solid vertical line in (a) and (b). (e) Normalized doping dependence of $\Delta \chi^{\prime\prime} (\omega, 104K,295K)=\chi^{\prime\prime}(\omega,104K)-\chi^{\prime\prime}(\omega,295K)$. (f) and (g) fits of the UD$<$2K Bi-2212 Raman responses at 295 K and 104 K. The spin singlet and the quasiparticle excitation are respectively described by the dashed-dotted line and the dashed line.}
\label{fig4}
\end{figure*}

We propose a simple fit to highlight the balance effect between the SSE and the QSPE peak with doping. The slight asymmetry of the SSE peak line shape mentioned above will not be taken into account in this fit. We describe the SSE peak by the imaginary part of the linear response function of an harmonic oscillator with natural frequency $2\Delta_{sse}$ and damping $\gamma_{sse}$ \cite{Hayes1978}. 
\begin{align}
	\chi^{\prime\prime}_{sse}(\omega)= \frac{C_{sse} \omega}{\gamma^2_{sse} \omega^2+(\omega^2-(2\Delta_{sse})^2)^2}
\end {align}
where $C_{sse} \propto \gamma_{sse}\Delta_{sse}$.
The QSPE peak is described by the imaginary part of the electronic polarization \cite{Klein1982}(often called the Drude response) and controlled by the damping $\gamma_{qsp}$: 
\begin{align}
	\chi^{\prime\prime}_{qsp}(\omega)= \frac{C_{qsp}\,\omega}{\gamma^2_{qsp} +\omega^2}
\end {align}
where $C_{qsp} \propto \gamma_{qsp}\eta_F$. $\eta_F$ is the density of state at the Fermi energy.
The Raman response $\chi^{\prime\prime}(\omega)$ is the sum of these two contributions. Selected fits are shown in Fig.~\ref{fig1} (b) to (e). They reproduce pretty well the experimental Raman spectra for several doping levels. We clearly see that the QSPE contribution (dashed line) forms gradually a hump (around 1000 \cm) and then a narrow peak (below 500 \cm) as the doping increases from the UD$>$15K to the OD63 K Bi-2212 compound. This is consistent with low energy Raman response calculation based on 2D Hubbard model \cite{Gull2013a}.  Conversely, the SSE peak (dashed-doted line) shifts to low energy, broadens and weakens in intensity. Our fits are in good agreement with the earlier computations \cite{Prelov1996} and in particular the Raman response calculation from the two dimensional Hubbard model using cluster dynamical mean field theory \cite{Lin2012}. In this calculation, the total diagrammatic bubble part and the vertex correction correspond respectively to the QSPE and SSE contribution. We interpret the growing up of the low energy QSPE peak  such as the release of mobile charges when the AF order is damaged by doping. The respective lifetimes of the spin singlet and quasiparticle excitations extracted from the fits are reported in Fig.~\ref{fig3}. They evolve in the opposite way with doping. At very low doping ($p \leq 0.06$), the lifetime of the spin singlet excitation is larger than the one of the quasiparticles. Beyond $p=0.15$, the QSP  lifetime is larger than the SSE one. Interestingly, it exponentially increases when approaching $p=0.22$. This specific doping was considered as the ending point of the pseudogap phase in Bi-2212 cuprate \cite{benhabib15,Loret2017a} where changes in Fermi surface topology have been detected \cite{benhabib15,Fang2022}.  
%%%%%%%%%%%%%%%%%%%%%%%%%%%%%%%%%%%%%%%%%%%%%%%%%%%%%%%%%%%%%%%%%%%%%%%%%%%%%%%%%%%%%%%%%%%%%%%%%%%%%%%%%%%%%%%%%%
%%%%%%%%%%%%%%%%%%%%%%%%%%%%%%%%%%%%%%%%%%%%%%%%%%%%%%%%%%%%%%%%%%%%%%%%%%%%%%%%%%%%%%%%%%%%%%%%%%%%%%%%
We now address the question of the pseudogap signature with temperature \cite{Alloul89,Norman1998,Ishida1998, Tallon01,Venturini2002a,Kohsaka2008}. In Fig.~\ref{fig4} (a)-(c), are displayed the Raman responses of the UD$>$2K, UD$<$15K and UD 50K Bi-2212 compounds above and below \Ts (T= 295 K and 104 K respectively). The pseudogap signature is highlighted by the blue shaded zone which circumscribes the loss of low energy spectral weight in the Raman response when temperature is lowered from T=295 K down to 104 K. We see that this loss decreases with doping. The low energy Raman spectral weight at $p=0.22$ (OD62K) exhibits a gain instead of a loss (indicated by a pink shaded zone in the inset of Fig.~\ref{fig4} (d)). The doping dependence of the subtracted Raman response, $\Delta \chi^{\prime\prime} (\omega, 104K,295K)=\chi^{\prime\prime}(\omega,104K)-\chi^{\prime\prime}(\omega,295K)$ is plotted in Fig.~\ref{fig4} (e). The crossing from spectral weight loss to weight gain is abrupt and indicated by an arrow nearby $p=0.22$ where the pseudogap collapses.  
%%%%%%%%%%%%%%%%%%%%%%%%%%%%%%%%%%%%%%%%%%%%%%%%%%%%%%%%%%%%%%%%%%%%%%%%%%%%%%%%%%%%%%%%%%%%%%%%%%%%%%%%%%%%%%%%%%
%%%%%%%%%%%%%%%%%%%%%%%%%%%%%%%%%%%%%%%%%%%%%%%%%%%%%%%%%%%%%%%%%%%%%%%%%%%%%%%%%%%%%%%%%%%%%%%%%%%%%%%%%%%%
Let's focus on the strongly under-doped UD$<2$K Bi-2212 compound (Fig.~\ref{fig4} (a)) where the pseudogap signature is strongest. We propose to interpret the pseudogap depletion by considering the balance effect between the quasiparticles and spin singlet excitations as a function of temperature (instead of doping as discussed above). Fits of the Raman responses measured at T= 295 K and 104 K are reported in Fig.~\ref{fig4} (f) and (g). These fits highlight the temperature dependence of the QSPE and SSE peak. As the temperature decreases, the QSPE peak contribution (dashed line) is strongly reduced at low energy (below 2000 \cm) while the SSE peak contribution (dash-dotted line) at higher energy becomes stronger and narrower around 3000 \cm. The conjunction of these two phenomena produces the loss of low energy spectral weight in the Raman response as pointed out by the blue shaded zone in Fig.~\ref{fig4} (a). The fits parameters are listed in Appendix C. The pseudogap is then interpreted by considering the enhancement of the  AF correlations upon cooling which accentuates the blocking of the mobile charge carriers. The residual magnetic correlations are likely short-range in space and dynamical in time and could be responsible for the pseudogap as suggested theoretically before \cite{Kyung2006,Lin2012,Wu2018}. Our experimental observations and interpretation are compatible with the temperature dependence of the \BAN Raman response at low doping calculated for the 2D Hubbard model \cite{Lin2012}.On the other hand, in the strongly over-doped regime, beyond the ending point of the pseudogap (OD62K Bi-2212, Fig.~\ref{fig4} (d)), the low energy QSPE becomes narrower and more intensive as the temperature decreases, giving rise to a well defined low energy peak as expected for the metallic Raman response \cite{Devereaux2007} and predicted by 2D Hubbard model computation \cite{Gull2013a,Lin2012}. 

So far, we have studied the balance effect between the low energy QSPE and the high energy SSE above $T_{c}$. Let's focus now on the relationship between the Bogoliubov quasiparticles excitation and the SSE.  In Fig.~\ref{fig5} (a), are displayed the ($B_{1g}+A_{2g}$) Raman responses of Bi-2212 single crystals over an extended range of doping measured at low temperature (T=15 K). 
%%%%%%%%%%%%%%%%%%%%%%%%%%%%%%%%%%%%%%%%%%%%%%%%%%%%%%%%%%%%%%%%%%%%%%%%%%%%%%%%%%%%%%%%%%%%%%%%%%%%%%%%%%%%%%%%%%%%%%%%%%%%%%%%%%%%%%%%
\begin{figure*}[ht]
\includegraphics[scale=0.65]{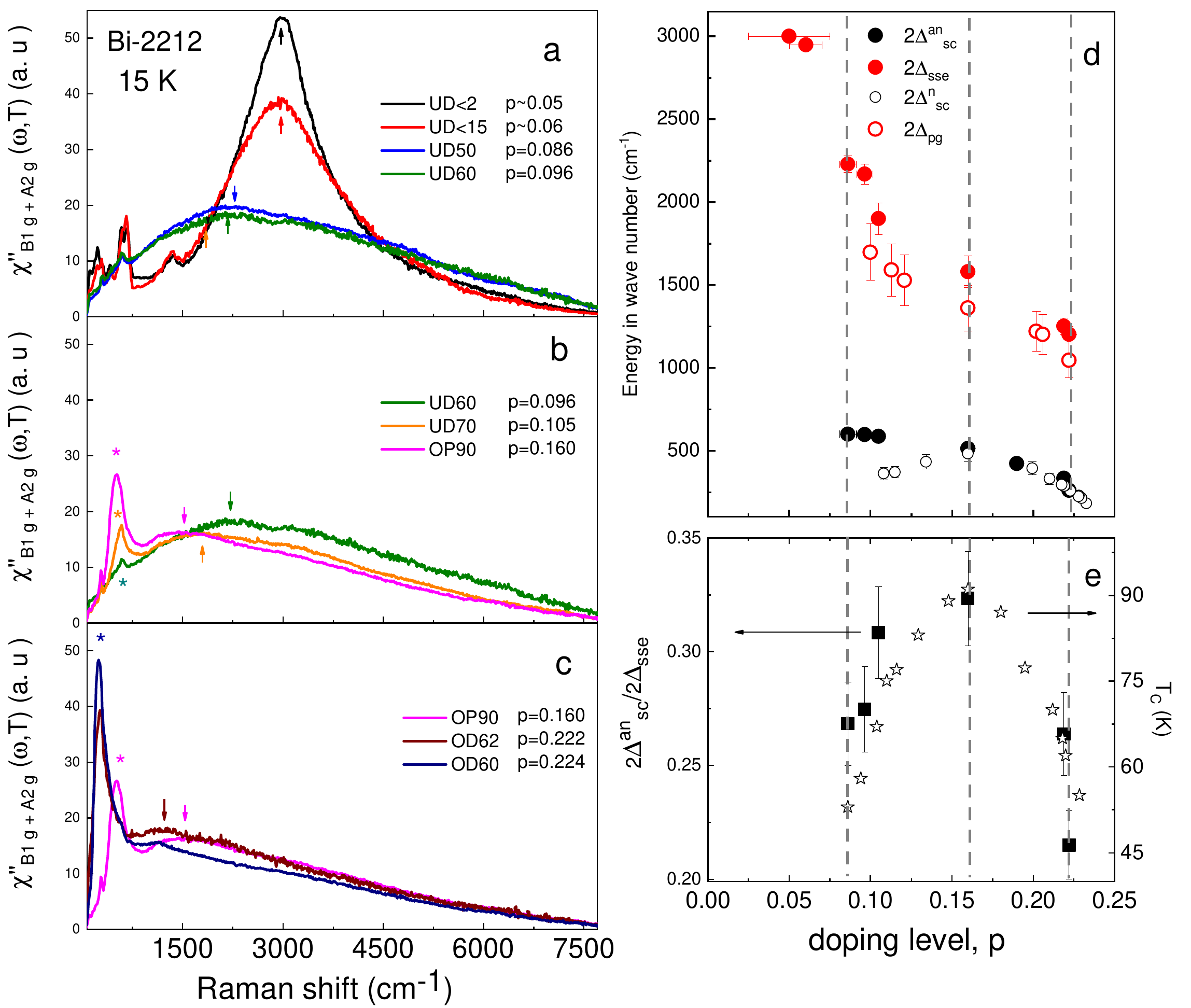}
\caption{(a)-(c) $\chi^{\prime \prime}_{B1g+A2g} (\omega)$ of Bi-2212 at $T=15$ K for selected doping level. The arrows indicate the locations of SSE peaks and the stars the locations of the pair breaking peaks. (d) Energy scale of the spin single excitations $2\Delta
_{SSE}$, the  anti-nodal SC gap $2\Delta^{\textrm{an}}_{\textrm{sc}}$ and the nodal SC gap $2\Delta^{n}_{\textrm{sc}}$ extracted from our previous works \cite{Loret2020}. (e)  Doping dependence of the $\frac{2\Delta^{\textrm{an}}_{\textrm{sc}}}{2\Delta_{\textrm{sse}}}$ ratio (black squares) which follows \Tc (open stars).The dotted lines are guides for the eyes.}
\label{fig5}
\end{figure*}
%%%%%%%%%%%%%%%%%%%%%%%%%%%%%%%%%%%%%%%%%%%%%%%%%%%%%%%%%%%%%%%%%%%%%%%%%%%%%%%%%%%%%%%%%%%%%%%%%%%%%%%%%%%%%%%%%%%%%%%%%%%%%%%%%%%%%%%%
At low doping ($p=0.05$ and $p=0.06$), the spin singlet excitation is still there located at $3000$ \cm (indicated by arrows). The narrow features below 1500 \cm are phonon modes. The weak one at 1300 \cm is a double phonon of the stronger one at 650 \cm. These modes weaken with doping due to their scattering with the increased number of carriers. As the doping increases, we see that the SSE peak broadens to turn into a bump and its maximum shifts to lower energy (2250 \cm for $p \approx 0.1$) (cf. Fig.~\ref{fig5} (b)). For $p>0.1$ (cf. Fig.~\ref{fig5} (b-c)), the SSE hump softens in energy down to 1300 \cm and fades away. The maximum of the hump disappears beyond $p=0.22$ (see the dark blue curve of the OD60K Bi-2212 compound in Fig.~\ref{fig5} (c)).  The weak feature observed around 1300 \cm is a double phonon already mentioned above. The superconductivity appears clearly in the ($B_{1g}+A_{2g}$) spectra above p$\geq 0.10$. It manifests itself by a Cooper pairs breaking peak associated with the Bogoliubov quasiparticles excitation \cite{Venturini2002,Hewitt2002,Masui2003,Blanc2009, Munnikes2011,benhabib15}. It is labeled by a star in Fig.~\ref{fig5} (b-c). In this geometry, the pair breaking peak corresponds to the SC gap around its maximum amplitude i.e. the anti-nodal (an) SC gap. The pair breaking peak is getting stronger with doping (Fig.~\ref{fig5} (b-c)). The SC pair breaking peak is associated with a dip on its higher energy side as it can be seen in the OP90 Raman response (pink curve in Fig.~\ref{fig5}(b)). This originates from the interplay between the superconducting gap and the pseudogap in the anti-nodal region as shown in our previous works \cite{Loret2016,Loret2017a,Loret2018}. The peak dip hump structure is also reproduced from other models \cite{Chubukov1999,Gull2013a}.

The energy of the SSE, $2\Delta_{sse} (p)$ (red filled circles) and the one of the pseudogap, $2\Delta_{pg}(p)$ (open red circle) are shown in in Fig.~\ref{fig5} (d). The $2\Delta_{pg} (p)$ plot was taken from the ref.\cite{Loret2020}. Its energy was defined as the energy of the dip-end (mentioned just above) and detected in the superconducting Raman response \cite{Loret2020}.  We see that both $2\Delta_{sse}(p)$  and $2\Delta_{pg}(p)$ decrease with doping and they follow each other closely in the doping range where they were both measured. Moreover, they both disappear at the same doping level $p=0.22$ where the pseudogap collapses \cite{benhabib15,Loret2017a}. This experimental observation lead us to conclude that $2\Delta_{pg} (p)$ and  $2\Delta_{sse}$ corresponds to the same energy scale. 

The energy of the antinodal  (an) SC gap, $2\Delta^{\textrm{an}}_{\textrm{sc}}$ (black filled circles) and the one of the nodal (n) SC gap, $2\Delta^{\textrm{n}}_{\textrm{sc}}$ (open black circles) versus doping, are displayed in Fig.~\ref{fig5} (d).  The energy of the (n) SC gap versus doping was taken from the ref.\cite{Loret2020}. The (n) SC gap is detected in the \BN geometry and corresponds to the amplitude of the SC gap around the nodes. The energy of the (n) SC gap follows a dome like shape approximately similar to that of the critical temperature \Tc as a function of doping (described by open stars in Fig.~\ref{fig5} (e)). On the contrary the energy of the (an) SC gap does not follows \Tc and depart from the (n) SC gap in the underdoped regime. The distinct behaviour between the (n) and (an) SC gap with doping has long been established by Raman scattering \cite{LeTacon2006,Devereaux2007,Guyard2008,Munnikes2011} and raised many questions (still debated) among which the existence or not of a relationship between the (an) SC gap and \Tc as it is the case for the (n) SC gap. 

In an attempt to clarify this last point, we focus first on the possible link between the energy of the SSE and the one of the anti-nodal SC gap. We see in Fig.~\ref{fig5} (d) that the energy of the SSE and the anti-nodal SC gap exhibit the same trend i.e: a decreasing with doping. However, they are not proportional to each other with doping as suggested in earlier studies \cite{Sugai2000,Li2012,Wang2022}. We find that the $\frac{2\Delta^{\textrm{an}}_{\textrm{sc}}}{2\Delta_{\textrm{sse}}}$ ratio versus doping (black square in Fig.~\ref{fig5} (e)) is not a constant. It rather describes a dome like shape with a maximum close to the optimal doping $p=0.16$ exactly like \Tc, plotted as open stars in Fig.~\ref{fig5} (e).
This lead us to  propose the following empirical relation: 
\begin{align}
2\Delta^{\textrm{an}}_{\textrm{sc}}(p) \propto 2\Delta_{\textrm{sse}}(p) \, \Tc(p)
\end{align}
Since, we have defined $2\Delta_{\textrm{sse}}(p)= J_{\textrm{eff}}(p)$, we get: 
\begin{align}
2\Delta^{\textrm{an}}_{\textrm{sc}}(p) \propto \Delta_{\textrm{sc}}^{\textrm{max}}(p) \propto {J_{\textrm{eff}}(p)}\,  \Tc(p)  \label{eqscale}
\end{align}

$J_{\textrm{eff}}(p)$ links $2\Delta^{an}_{sc} (p)$ to \Tc. This does not allow us to specify explicitly the role played by the antiferromagnetic correlations in the SC paring mechanism without theoretical calculations;  even if, recent investigations carried out by other techniques support that Cooper pairing is driven by spin fluctuations in cuprates \cite{LeTacon11,Restrepo2022,Mahony2022b,Ma2022}. We hope that our empirical relation could be checked in a near future, by cluster dynamical mean field calculations of the Raman responses of the 2D Hubbard model in the superconducting state. This required computation of the Raman response over a large spectral range and large doping range, so that, the energies of $\Delta_{\textrm{sc}}^{\textrm{max}}(p)$ and $J_{\textrm{eff}}(p)$ could be estimated. 

Finally, we would like to stress that this empirical relation is consistent with previous ones got from different experimental techniques.  As seen before, $J_{\textrm{eff}}(p)$ is related to the energy of the Raman SSE peak. It characterizes the strength of residual antiferromagnetic correlations in the metallic phase state and their ability to block the mobile charge carriers as a function of doping, $p$. On the other hand, the mobile charge carriers manifests themselves by the Raman QSPE peak whose intensity is related to the quasiparticle spectral weight at the antinodes  $Z_{an}(p)$ \cite{Devereaux2007}.  As $J_{\textrm{eff}}(p)$ gets stronger, the number of charge carriers decreases and therefore
$Z_{an}(p)$ decreases. It is then reasonable to conjecture that $J_{\textrm{eff}}(p) \propto \frac{1}{f(Z_{an}(p))}$ where $f$ is a monotonic function.  The empirical relation \eqref{eqscale} becomes: $\Delta_{\textrm{sc}}^{\textrm{max}}(p)\, f(Z_{an}(p)) \propto \Tc(p)$. This relation is similar to the one earlier proposed in ref.\cite{Ding01} and based on angular resolved photo-emission spectroscopy measurements in underdoped Bi-2212 compounds. Indeed, they reported that the ratio $\frac{\Delta_{\textrm{sc}}^{\textrm{max}} (p).Z^{arpes}_{an} (p)}{\Tc}$ is approximatively a constant with doping. $Z^{arpes}_{an}$ is the coherent quasiparticle spectral weight extracted from the spectral function at the antinodes.

More recently the combination of the Homes'law \cite{Homes2005} in the dirty limit ($\sigma_{dc}\Delta_{\textrm{max}} \propto n_s$) and the Uemura law \cite{Uemura1989} valid in the underdoped regime: ($n_s\propto \Tc$) lead to ($\sigma_{dc} \, \Delta_{\textrm{sc}}^{\textrm{max}} \propto \Tc$) where $n_s$ and $\sigma_{dc}$ are respectively the superfluid density and the dc-conductivity. This equation is also consistent with our empirical relation \eqref{eqscale}, provided that  $\sigma_{dc}(p) \propto \frac{1}{J_{\textrm{eff}}(p)}$. This is understandable, if we consider that the stronger $J_{\textrm{eff}}(p)$ is, the more difficult it will be to move the carriers. 

\section{III. Conclusion}

In conclusion, by tracking simultaneously the doping dependence of the spin singlet and the quasiparticle excitations respectively related to the AF lattice and the mobile charge carriers, we can follow step by step the transition of the Bi-2212 cuprate from its Mott insulating antiferromagnetic phase to its metallic phase. As the AFM order is damaged by doping, the mobile charge carriers are released and the QSPE lifetime exponentially increases when approaching the pseudogap end. We interpret the pseudogap effect with temperature as the blocking of mobile charge carriers by the enhancement of the AF magnetic correlations as the temperature is lowering. The pseudogap energy scale, $\Delta_{\textrm{pg}}(p)$ closely follows the one of the spin singlet excitation, $\Delta_{\textrm{sse}}(p)$. In the superconducting state, we show that $2\Delta^{\textrm{an}}_{sc}(p) \propto {J_{\textrm{eff}}(p)} \, \Tc(p)$. This empirical relation establishes an intriguing link between the (an) SC gap and \Tc and suggests that the effective exchange energy should play a key role in the SC mechanism. 

\par

\section {Acknowledgments} We are grateful to L. Taillefer, A.Georges, A. Millis, I. Paul, M. Civelli, Y. Sidis and M.H.Julien and C.Proust for useful discussions. 
We acknowledge support from  the ANR grant NEPTUN n◦ANR-19-CE30-0019-01, the CNRS National Institute of Physics and the Mission for Transversal and Interdisciplinary Initiatives (MITI) of the CNRS. Work at Brookhaven is supported by the Office of Basic Energy Sciences, Division of Materials Sciences and Engineering, U.S. Department of Energy under Contract No. DE-SC00112704. 

%\section{Author contributions}
%M.M. performed the Raman measurements with the help of A.A., A.S., Y.G.,M.C. and M.M and A.S. performed the data analysis. R.D.Z, J.S., G.D.G  grew the single crystals.  R.D.Z, J.S., G.D.G and D.C. performed the annealing procedure to obtained underdoped and overdoped composition. G.D.G.,R.D.Z, J.S., D.C and M.M. characterized the samples.  A.S. wrote the manuscript with the input from all the authors. A.S. supervised the project.

%\section{Competing interests}
%The authors declare no competing interests

%\section{Materials and Correspondence}
%Correspondence and request for materials should be addressed to A.S. (alain.sacuto@univ-paris-diderot.fr). 

%\textbf{Data availability}
%All data generated or analyzed during this study are included in the published manuscript.

\appendix
\section{Appendix A: Details of the Raman experiments} 

Raman experiments were carried out using a JY-T64000 spectrometer in a single grating configuration using a 600 grooves/mm grating and  Thorlabs notch filters to block the stray light. The spectrometer is equipped with a nitrogen cooled back illuminated 2048x512 CCD detector. We use the 532 nm and 488 nm excitation lines. Measurements between 10 and 295 K have been performed using an ARS closed-cycle He cryostat. This configuration allows us to cover a wide spectral range ($70~cm^{-1}$ to $8000~cm^{-1}$) with a resolution sets at $9~cm^{-1}$. Each spectrum has been obtained from several frames  centred on different wavelengths to cover the whole spectral range. Each frame is repeated twice to eliminate cosmic spikes and acquisition time is about 10 mn. All the spectra have been corrected for the Bose factor and the instrumental spectral response. They are thus proportional to  the Raman response function $\chi^{\prime \prime}(\omega,T)$. The $B_{1g}+A_{2g}$ and $B_{2g}+A_{2g}$ Raman responses have been obtained from cross polarizations at 45$^o$ from the Cu-O bond directions and along them respectively.

\section{Appendix B: Doping evolution of the SSE peak from strongly to moderate underdoping} 
\begin{figure}[ht!]
\includegraphics[scale=0.38]{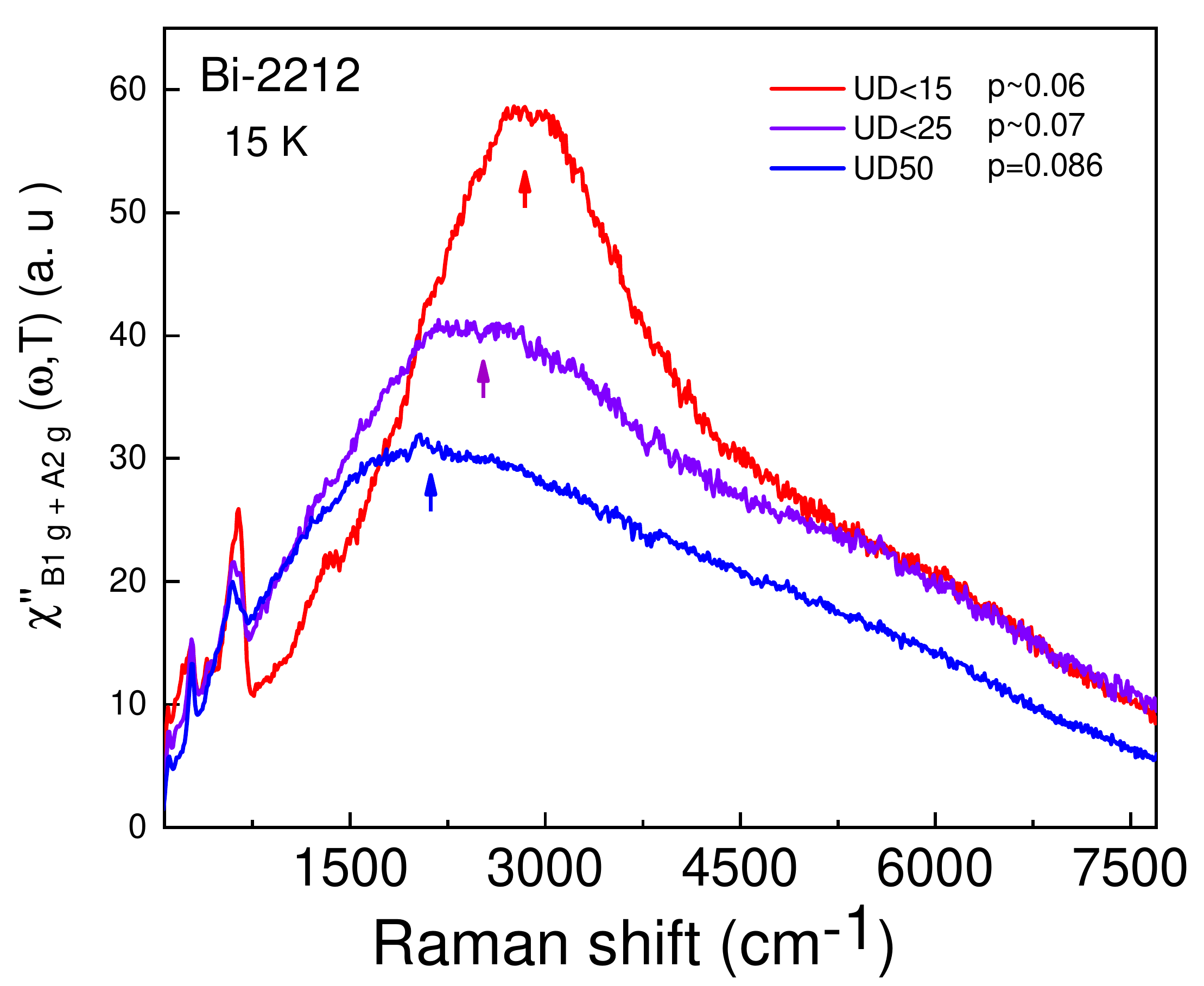}
\caption{(a) $\chi^{\prime \prime}_{B1g+A2g} (\omega)$ of Bi-2212 compound as a function of doping measured à T=$15$ K. The 488 nm excitation line was used.}
\label{fig6}
\end{figure}
We can clearly track the  SSE peak position as a function of doping level. It shifts to low energy and it broadens as the doping increases (see the arrows in Fig.~\ref{fig6}).  
The SSE peaks intensity is approximately 1.5 higher than the one measured  with the 532 excitation line and its high energy asymmetric part is sightly enhanced. This is consistent with earlier Raman studies \cite{Lyons1988,Sugai1988,Blumberg1997, Chelwani2018}.

\section{Appendix C: Fits parameters of the the SSE and QSPE peaks in the Raman response} 
The fits parameters of the UD$<$2K Bi-2212 Raman responses are : $\gamma_{sse}=1400$ \cm, $\gamma_{qsp}=2000$ \cm  and $2\Delta_{sse}=2800 \, \cm$ for $T=295$ K and $\gamma_{sse}=1200$ \cm , $\gamma_{qsp}=6000$ \cm and $2\Delta_{sse}=3000$ \cm for $T=104$ K. In these fits, we made the choice to increase the $\gamma_{qsp}$ scattering rate rather than decreases the QSPE peak intensity, both options are possible. The fits parameters of the  OD62K Bi-2212 Raman responses are:  $\gamma_{qsp}=400$ \cm, $\gamma_{sse}=4800$ \cm and $\Delta_{sse}=2600 \cm$ for $T=295$ K and $\gamma_{qsp}=210$ \cm , $\gamma_{sse}=4600$ \cm and $\Delta_{sse}=2500 \cm$ for $T=104$ K.

\clearpage

* electronic address: alain.sacuto@univ-paris-diderot.fr
\bibliography{cuprates}

\end{document}